\documentstyle[12pt,epsfig,amsmath,amssymb,psfrag,pifont,cite]{article}
\numberwithin{equation}{section}
\numberwithin{table}{section}
\renewcommand{\d}{\partial}

\def\ga{\mathrel{\raise.3ex\hbox{$>$\kern-.75em\lower1ex\hbox{$\sim$}}}}
\def\la{\mathrel{\raise.3ex\hbox{$<$\kern-.75em\lower1ex\hbox{$\sim$}}}}

\def\I_M{{I_{\scriptscriptstyle M\times M}}}

\def\be{\begin{equation}}
\def\ee{\end{equation}}
\def\bea{\begin{eqnarray}}
\def\eea{\end{eqnarray}}

\newcommand{\beqal}{\begin{eqnarray}\label}
\newcommand{\beqa}{\begin{eqnarray}}
\newcommand{\eeqa}{\end{eqnarray}}

\begin{document}

\begin{titlepage}
\begin{center}

\vskip .2in

{\Large \bf Holographic electrical and thermal conductivity in strongly coupled gauge theory with multiple chemical potentials}
\vskip .5in

{\bf Sachin Jain\footnote{e-mail: sachjain@iopb.res.in}\\
\vskip .1in
{\em Institute of Physics,\\
Bhubaneswar 751~005, India.}}
\end{center}\noindent
\baselineskip 15pt

\begin{center} {\bf ABSTRACT}

\end{center}
\begin{quotation}\noindent
\baselineskip 15pt

We study transport coefficients of strongly coupled gauge theory in the presence of multiple chemical potential which are dual to rotating D3, M2 and M5 brane. Using the general form of the perturbation equations, we compute electrical conductivity  at finite temperature as well as at zero temperature.  We also study thermal conductivity for the same class of black holes and show that thermal conductivity and viscosity obeys Wiedemann-Franz like law even in the presence of multiple chemical potential. 
\end{quotation}
\vskip 2in
Dec 2009\\
\end{titlepage}
\vfill
\eject

\setcounter{footnote}{0}
\section{Introduction}
One of the hallmark of gauge/gravity duality is to allow us to 
gain insights into strongly coupled gauge theories both at zero and non-zero
temperature. Even though these theories, in several ways, different from
realistic gauge theories like QCD, they do share qualitatively similar behavior 
in many respects. For example, 
in large $N_c$ limit, and in the presence of an infrared cutoff, these theories 
de-confine at high temperature. Viscosity to entropy ratio of these theories 
are quite low as suggested by RHIC data. In all these cases, one could carry 
out trustable calculations using conjectured gauge-gravity duality. It allows us to map a strongly 
coupled gauge theories to a weakly coupled gravity. In particular, in the large $N_c$ and 
large t'Hooft coupling limit, the dual description turns out to be various 
black holes in anti deSitter (AdS) space. 

In the above framework  the low energy or long wavelength behavior
of ${\cal N} = 4$ supersymmetric $SU(N_c)$ Yang-Mills was studied in \cite{Son:2006em}, in the presence 
of non-zero chemical potential. Generically, three independent chemical potentials 
can be associated for three $U(1)$ charges in the Cartan sub-algebra of 
$SO(6)_R$. The dual then represents a five dimensional R-charged black hole as discussed in\cite{Behrndt:1998jd}. Though shear viscosity was calculated in the presence of three chemical potentials, thermal conductivity was calculated only when a single chemical potential is non-zero. It was found that an appropriate ratio of thermal 
conductivity and the 
shear viscosity is $8 \pi^2$. This is the analogous Wiedemann-Franz law for this 
gauge theory\cite{Son:2006em}. Subsequently there were many other works \cite{Mas:2006dy} (see\cite{Son:2002sd} and references therein) along these directions.    In a recent paper \cite{Jain:2009uj}, some of the above results 
were generalized in the presence of multiple chemical potential. Besides providing a general framework and solutions of certain long 
wavelength perturbation of the geometry, we computed diffusion coefficients 
associated with these black holes carrying multiple R-charges in four, five and seven 
dimensions. We found that near the second order unstable points of these holes, 
the diffusion coefficient is zero - a fact consistent with the critical slowing 
down\cite{Maeda:2008hn} . 

It turns out that, regardless of the detail structure of these 
black holes,  many of  the above calculations for electrical conductivity can be simplified and can be given a 
general structure. This is what we first attempt to do in this paper. We then use 
these results to compute  
thermal conductivities in the presence of {\it multiple} R-charges in {\it various} 
dimensions. Furthermore, we show that the thermal conductivity and viscosity ratio 
remains same and is equal to their single charge cosines. Subsequently, we turn our 
attention to the extremal R-charged black holes. Recently, it was understood that
studying 
perturbations of these black holes requires special care \cite{Faulkner:2009wj}. The approach was then 
used to compute viscosity to entropy for a rather large class of extremal black holes\cite{Edalati:2009bi}. In this paper, we also compute the conductivity of gauge theory dual to multiple R-charged
extremal  black hole and give a general expression for the dimension of dual operators. We then find an effective schrodinger problem 
which sheds light to some of the results obtained in the text. 

This paper is structured as follows. In the next section, we give a general framework 
to study relevant perturbation equations in the gravity side. This simplifies the computations
of previous works. In section 3, we use effective action approach to give a general 
expression of the electrical conductivity matrix in a way which is insensitive to 
many details of the explicit geometry. In section 4, we carry out an analysis and
discussion on 
conductivity for black holes with different charges individually. In the later section, we turn our attention 
on the computation of thermal conductivity for multiple R-charged black 
holes and analyze Wiedemann-Franz law. We find that in five, four and seven 
dimensions the appropriate ratio of thermal conductivity and viscosity, regardless
of 
the number of charge contents,  are $8 \pi^2$, $32 \pi^2$ and $2 \pi^2$ 
respectively. In section 6, we study the electrical conductivity of the 
extremal R-charged black holes. In the next section, we map the problem to Schr\"{o}edinger problem and find out behavior of effective Schr\"{o}edinger potential for both extremal as well as non-extremal case. Last section of this paper contains a discussion of our result. 
At the end we collect some earlier results which are used in the
main body of the paper.

\section{General perturbation equation}
In this section we write down perturbation equation required for computation of electrical conductivity. For that we consider a lagrangian of the form
\begin{equation}
 {{\cal L}\over
 \sqrt{-g} }=
 R  - {1\over 4} G_{ij} F_{\mu\nu}^i 
 F^{\mu \nu\, j} + \rm{Scalar~ field~ terms}
+ \rm{Chern-Simons~ term},
\label{lagrangian}
\end{equation}
where $ F^{\mu \nu\, j}$ is the field-strength tensor of the $j$-th $ U(1) $ gauge field. The metric that we take is of the form
\begin{equation}
 ds^{2} = g_{tt} dt^{2} + g_{uu} du^{2} + g_{xx} \sum_{i=1}^{d-1} (dx^{i})^{2}.
\end{equation}
 As an example one can consider R-charged black holes in various dimensions (see \cite{Son:2006em},\cite{Behrndt:1998jd}).

Since our aim is to compute the electrical conductivity using Kubo formula, it is sufficient to consider perturbations in the tensor (metric)  and the vector (gauge fields) modes around the black hole solution and keep other fields such as scalars unperturbed. So perturbations are of the form: 
\begin{equation}
g_{\mu\nu}= {\bf g}^{(0)}_{\mu\nu} + h_{\mu\nu}~,
\quad\quad
A^{i}_\mu = {\bf A}^{i(0)}_\mu + {\cal A}^i_\mu~,
\quad i = 1,2\quad.
\end{equation}
where ${\bf g}^{(0)}_{\mu\nu}$ and ${\bf A}^{i(0)}_\mu $ are background metric and gauge fields.
In order to determine R-charge conductivity it is enough to  consider perturbations in $(tx)$ and 
$(xx)$ component of the metric tensor and $x$ component of the gauge fields. Moreover one can 
choose the perturbations to depend on radial coordinate $u$, time $t$ and one of the spatial 
worldvolume coordinate $z$.
A convenient ansatz with the above restrictions in mind is
\begin{equation}
h_{tx} = {\bf g}_{0xx}~T(u)~e^{-i\omega t + i q z},\quad
h_{zx} = {\bf g}_{0xx}~Z(u)~e^{-i\omega t + i q z},\quad
{\cal A}^i_x = \frac{\mu^{i}}{2} ~ \Phi_{i}(u)~~e^{-i\omega t + i q z}.\quad
\end{equation}
Here $\omega$ and $q$ represent the frequency and momentum in $z$ direction respectively and we set perturbations in the other components to be equal to zero. Also note that, $\mu^i$ represents the chemical potential for the $i^{th}$ gauge field.                                                                                Next step is to find linearized equations which follows from the equations of motion. It turns out that at the level of linearized equation and at zero momentum limit metric perturbation $Z(u)$ decouple from the rest. One can further eliminate $T(u)$ reducing it to equation for perturbations in gauge fields only. After substitution one finds the equations for perturbed gauge fields to be
\begin{eqnarray}
 \frac{d^2}{du^2}\Phi_i(u)&+&\Big(\frac{d}{du}\log(\sqrt{-g}G_{ii}g^{xx}g^{uu})\Big)\frac{d}{du}\Phi_{i}(u)-\omega^2\frac{g_{uu}}{g_{tt}}\Phi_{i}(u)\nonumber\\
&+&\frac{1}{g_{tt}}\frac{F_{ut}^i}{\mu^i}\Big(\sum\limits_{j=1}^m \mu^j G_{jj}F_{ut}^j\Phi_{j}(u)\Big)=0.\label{generalwithchemical}
\end{eqnarray}
 The last term in equation (\ref{generalwithchemical}) comes from the substitution of the metric perturbation part which can be looked upon as interaction between different gauge fields. Note that all the equation written in \cite{Jain:2009uj},\cite{Maeda:2008hn}  can be reproduced from the equation (\ref{generalwithchemical}). One can rewrite equation (\ref{generalwithchemical}) as

\begin{equation}
\frac{d}{du}(N_i\frac{d}{du}\Phi_{i}(u))-\omega^2\ N_i~ g_{uu} g^{tt}\Phi_i(u)+\sum\limits_{j=1}^m  M^{1}_{ij}\Phi_{j}(u)=0,
\end{equation}   
where
\begin{equation}
N_i=\sqrt{-g}G_{ii}g^{xx}g^{uu}. 
\end{equation}
and 
\begin{equation}
 M^{1}_{ij}=\frac{F_{ut}^i}{\mu^i}\sqrt{-g}G_{ii}g^{xx}g^{uu}g^{tt}\mu^j G_{jj}F_{ut}^j.
\end{equation}
Note that $ M^{1}_{ij}\neq M^{1}_{ji}$.
As an example consider R-charge black holes in 5-dimension (with 3 R-charges, see \cite{Jain:2009uj}). For this case we get 
\begin{eqnarray}
&&(\Phi_i)^{\prime\prime} + ~\left(\dfrac{f^\prime}{f} + 2 \dfrac{H_i^\prime}{H_i} -  \dfrac{\cal H^\prime}{\cal H}  \right) (\Phi_i)^\prime
 + \dfrac{(\frac{\omega}{2 \pi T_{0}})^{2} {\cal H}}{f^2 u} \Phi_i 
- \dfrac{u(1+k_i)}{fH_i^2}
\Big[ k_1(1+k_2)(1+k_3) \Phi_1 \nonumber \\
&&+ k_2(1+k_1)(1+k_3)\Phi_2 +  k_3(1+k_1)(1+k_2)\Phi_3 \Big] = 0. \label{D3eq}
\end{eqnarray}

It is convenient to redefine $\Phi$ as $\phi_{j}(u)=\mu^j\Phi_{j}(u)$.\footnote{No summation. Unless explicitly written there is no summation for repeated indices in this paper.}  In the redefined variables equations are 
\begin{equation}
\frac{d}{du}(N_i\frac{d}{du}\phi_i(u))-\omega^2 N_i~ g_{uu} g^{tt} \phi_i(u)+\sum\limits_{j=1}^m  M_{ij}\phi_j(u)=0.
\label{eqnmotion}
\end{equation}
with
\begin{equation}
N_i=\sqrt{-g}G_{ii}g^{xx}g^{uu}. 
\end{equation}
and 
\begin{equation}
 M_{ij}=F_{ut}^i \sqrt{-g}G_{ii}g^{xx}g^{uu}g^{tt}G_{jj}F_{ut}^j.
\end{equation}
Note that $ M_{ij}= M_{ji}$.
For evaluating conductivity in the low frequency limit for non-extremal backgrounds we only need to solve equations up to zeroth order in $\omega$. To that order one finds,
 \begin{equation}
\frac{d}{du}(N_i\frac{d}{du}\phi_i(u))+\sum\limits_{j=1}^m M_{ij}\phi_j(u)=0.\label{eqnmotion1}
 \end{equation}
\section{Effective action and expression for conductivity}
In this section, after writing down effective action (which reproduces (\ref{eqnmotion1})), we extract out the expression for electrical conductivity using Kubo formula.
Effective action (for discussion on effective action for single charge case see \cite{Myers:2009ij}) can be written as
\begin{eqnarray}
S &=& \frac {1}{16\pi G_{d+1}} \int \frac{d^{d}q}{(2\pi)^d}du \Big[\sum\limits_{i=1}^m N_i(u)\frac{d}{du}\phi_{i}(u,\omega,q)\frac{d}{du}\phi_{i}(u,-\omega,-q)\nonumber \\
&+&\sum\limits_{i,j=1}^m M_{ij}(u)\phi_{i}(u,\omega,q)\phi_{j}(u,-\omega,-q)  \Big].
\end{eqnarray}
Boundary action is given by
\begin{equation}
S_\epsilon =\lim_{u\rightarrow 0}\frac {1}{16\pi G_{d+1}} \int \frac{d^{d}q}{(2\pi)^d}\left(\sum\limits_{i=1}^m N_i(u)\frac{d}{du}\phi_{i}(u,\omega,q)\phi_{i}(u,-\omega,-q)\right).
\end{equation}
For calculating conductivity we need to look for 
\begin{equation}
 \Im\Big[\lim_{u\rightarrow 0}\frac {1}{16\pi G_{d+1}} \int \frac{d^{d}q}{(2\pi)^d}\left(\sum\limits_{i=1}^m N_i(u)\frac{d}{du}\phi_{i}(u,\omega,q)\phi_{i}(u,-\omega,-q)\right)\Big]
\label{bdyaction}
\end{equation}
Now
\begin{eqnarray}
&&\frac{d}{du}\Im \Big(\sum_{i=1}^m N_i(u)\frac{d}{du}\phi_{i}(u,\omega,q)\phi_{i}(u,-\omega,-q)\Big)\nonumber\\
&&=\Im \Big[\sum_{i=1}^m \frac{d}{du}(N_i(u)\frac{d}{du}\phi_{i}(u,\omega,q))\phi_{i}(u,-\omega,-q)\nonumber\\
&&+\sum_{i=1}^m N_i(u)\frac{d}{du}\phi_{i}(u,\omega,q)\frac{d}{du}\phi_{i}(u,-\omega,-q)\Big].
\end{eqnarray}
Using (\ref{eqnmotion1}) r.h.s of above equation reduces to
\begin{equation}
\Im\Big[-\sum\limits_{i,j=1}^m M_{ij}(u)\phi_{i}(u,\omega,q)\phi^j(u,-\omega,-q)+\sum\limits_{i=1}^m N_i(u)\frac{d}{du}\phi_{i}(u,\omega,q)\frac{d}{du}\phi_{i}(u,-\omega,-q)\Big]  ,
\end{equation}
which is equal to zero since quantity in the bracket is real.
 Then (\ref{bdyaction}) can as well be calculated at the horizon i.e. at $u=1$, which simplifies calculations significantly. 
Regularity at the horizon implies
\begin{equation}
\lim_{u\rightarrow 1}\frac{d}{du}\phi_i(u)=-i \omega \lim_{u\to 1}\sqrt{-\frac{g_{uu}}{g_{tt}}} \phi_{i}(u)+{\mathcal{O}}(\omega^{2}). 
\end{equation}
Hence (\ref{bdyaction}) reduces to 
\begin{equation}
\Im\Big[-i \omega\lim_{u\rightarrow 1}\frac {1}{16\pi G_{d+1}} \int \frac{d^{d}q}{(2\pi)^d}\sqrt{\frac{g_{uu}}{g_{tt}}}\left(\sum\limits_{i=1}^m N_i(u)\phi_{i}(u,\omega,q)\phi_{i}(u,-\omega,-q)\right)\Big].
\end{equation}
Finally we get boundary action to be
\begin{equation}
S_\epsilon=\int \frac{d^{d}q}{(2\pi)^d}(\phi^i)^{(0)}(\omega, q){\cal F}_{ik} (\omega,q)\, (\phi^k)^{(0)} (-\omega, -q),
\label{bdyaction1}
\end{equation}
where $(\phi_{i})^{(0)}$ is the boundary value of the field $\phi_{i}$. Next, the retarded correlators
are given by
\begin{equation}
   G^R   \;=\;
  \begin{cases}  - 2  {\cal F}_{ik} (\omega,q)\,,
                                                    & i=k , \\
           \noalign{\vskip 4pt}
           -  {\cal F}_{ik} (\omega,q)\,,
                                                    & i\neq k . 
         \end{cases}
   \label{f_fth}
\end{equation}
The expression for diagonal and off diagonal parts\footnote{The coefficient of $(\phi_{i})^{0}(\phi_{j})^{0}$.} of conductivity can be written as 
\begin{equation}
\sigma_{ii}=-\lim_{\omega\rightarrow 0}\frac{\Im G^R(\omega,q=0)}{\omega}=\frac {1}{8\pi G_{d+1}}\Bigg[\sqrt{\frac{g_{uu}}{-g_{tt}}}\frac{\sum\limits_{k =1}^m N_k(u)\phi_k(u,\omega,q)\phi_{k}(u,-\omega,-q)}{(\phi_{i})^{0}(\phi_{i})^{0}}\Bigg]_{u\to 1,q\to 0},
\end{equation}

\begin{equation}
\sigma_{ij}=-\lim_{\omega\rightarrow 0}\frac{\Im G^R(\omega,q=0)}{\omega}=\frac {1}{16\pi G_{d+1}}\Bigg[\sqrt{\frac{g_{uu}}{-g_{tt}}}\frac{\sum\limits_{k=1}^m N_k(u)\phi_{k}(u,\omega,q)\phi_{k}(u,-\omega,-q)}{(\phi_{i})^{0}(\phi_{j})^{0}}\Bigg]_{u\to 1,q\to 0}.
\end{equation}
respectively.

Now we are ready to compute the electrical conductivity. We first present computation for single charge black hole and show that we get the same result as obtained in \cite{Maeda:2008hn} where conductivity was obtained by solving up to first order in $\omega$ (note that here we only need to compute zeroth order in $\omega$.). Then we 
compute same for multiple charges in various dimensions.

\section{Electrical conductivity}
In this section we compute electrical conductivity for gauge theories dual to 4, 5, 7 dimensional R-charge black hole. We observe that behavior of conductivity with temperature is $\sigma \sim T^{d-3}$ for d dimensional dual gauge theory. For multi-charge black hole we get conductivity matrix whose off diagonal parts comes solely due to effective interaction \footnote{This can be interpreted as, turning on electric field for one type not only effects fields which are charged under that field, it also effects  fields which are charged under other gauge fields as well, which is possible only when there exist interaction.} between gauge fields. We first compute conductivity with single chemical potential and then turn to multiple chemical potential cases.
\subsection{Single charge black hole in various dimension}
For single charge black hole one finds
\begin{equation}
%\begin{split}
(\phi_1)^{\prime\prime} + ~\left(\dfrac{f^\prime}{f} +  \dfrac{H_1^\prime}{H_1} -\frac{c}{u} \right) (\phi_1)^\prime - \dfrac{a u^{b}(1+k_1)}{fH_1^2}
\Big( k_1 \phi_1 \Big) = 0. \label{general1charge}
%\end{split}
\end{equation} 
The expression for conductivity reduces to 
\begin{equation}
\sigma=\frac {1}{8\pi G_{d+1}}\Big[\sqrt{\frac{g_{uu}}{-g_{tt}}}\frac{ N(u)\phi(u,\omega,q)\phi(u,-\omega,-q)}{(\phi)^{0}(\phi)^{0}}\Bigg]_{u\to 1,q\to 0}.
\end{equation}
Let us define ${\cal K} = \Big[ \frac {1}{8\pi G_{d+1}}\sqrt{\frac{g_{uu}}{-g_{tt}}}N(u)\Big]_{u\to 1} $ and $ F(u)=\phi_{1}(u,\omega,q)\phi_{1}(u,-\omega,-q)$.
\begin{itemize}
 \item \textbf{D=4:} In this case one gets $c=0, a=1,b=2$ and $m=0$.
Relevant parts are
\begin{equation}
{\cal K} = \frac{N_c^{\frac{3}{2}}}{24\sqrt{2}\pi}(1+k)^{\frac{3}{2}}.
\end{equation}
\begin{equation}
\phi(u)=\phi^{0}\frac{1+\frac{2 k u}{3}}{1+k u}
\end{equation}
which implies
\begin{equation}
\sigma={\cal K}\frac{F(u=1)}{F(u=0)}=\frac{(3+2 k)^2N_c^{\frac{3}{2}} }{6^{3}\pi\sqrt{2(1+k)}}.
\end{equation}
\end{itemize}

We see that for three dimensional gauge theory, conductivity is independent of temperature. Now we can compare this result with the the result for $\mu = 0$, case.
\begin{equation}
\frac{\sigma_\mu}{\sigma_{\mu=0}}=\frac{(1+\frac{2 k}{3})^2}{\sqrt{1+k}}\geq 1.
\end{equation}

 Since there exist a critical line $k=\frac{3}{2}$~ \cite{Maeda:2008hn}, one can not make conductivity arbitrarily large. This discussion also holds true for rest of the cases with only difference in location of critical line. 
\begin{itemize}
 \item \textbf{D=5:} Here $c=0, a=1,b=1$ and $m=0$.
Summary of the results are
\begin{equation}
{\cal K} =\frac{N_c^{2}T_{0}(1+k)^{\frac{3}{2}}}{16 \pi},
\end{equation}
\begin{equation}
 \phi(u)=\phi_{0}\frac{1+\frac{ k u}{2}}{1+k u}.
\end{equation}
So one gets conductivity 
\begin{equation}
\sigma=\frac{(2+ k)^2 N_c^{2}T_{0} }{64\pi\sqrt{(1+k)}}=\frac{N_c^{2}T_{H}(2+k)}{32 \pi}.
\end{equation}
where $T_{H}=\frac{(2+k)T_{0}}{2\sqrt{1+k}}$ is the hawking temperature of the black hole.
\end{itemize}

\begin{itemize}
 \item \textbf{D=7:} In this case $c=-1,a=4,b=3$ and $m=1$.
Relevant parts are
\begin{equation}
{\cal K} =\frac{4 N_c^{3}T_{0}^{3}(1+k)^{\frac{3}{2}}}{81},
\end{equation}
\begin{equation}
 \phi(u)=\phi_{0}\frac{1+\frac{ k u^{2}}{3}}{1+k u^{2}}.
\end{equation}
Conductivity in this case is given by
\begin{equation}
\sigma=\frac{4 (3+ k)^2 N_c^{3}T_{0}^{3} }{3^{6}\sqrt{(1+k)}}=\frac{4 N_c^{3}T_{H}^{3}(1+k)}{27(3+k)}.
\end{equation}
where $T_{H}=\frac{(3+k)T_{0}}{3\sqrt{1+k}}$ is the hawking temperature of the black hole.
\end{itemize}
% \begin{itemize}
%  \item So we have shown that we get same results for single charge case as obtained in \cite{Maeda:2008hn} where authors solved up to first order in $\omega$ to obtain the results.
% \end{itemize}
% \begin{itemize}
%  \item It is also concluded that $\frac{\sigma_\mu}{\sigma_{\mu=0}}\geq 1$ and there exist a upper bound on conductivity determined by critical line.
% \end{itemize}
% \begin{itemize}
%  \item There exist general result that \cite{Iqbal:2008by}, $\sigma_{\mu=0}|_{u=1} = \sigma_{\mu=0}|_{u=0} $. In the presence of chemical potential one gets $ \sigma^{Horizon}_{\mu} > \sigma^{Boundary}_{\mu}$. This result can be interpreted in terms of effective Schr\"{o}edinger potential . For $ \mu =0 $ case , what we expect is to get a potential which is constant (zero) through out where as for $ \mu \neq 0 $ one expects  varying potential.
% \end{itemize}
% \begin{itemize}
%  \item Note that the parameter $ k $ can be expressed in terms of $\Omega(=\frac{\mu}{T_H})$. It can be shown that at the critical point, though conductivity remains nonzero, its derivative w.r.to $\Omega$ diverges (with the critical index $\frac{1}{2}$ for four dimensional gauge theory). 
% \end{itemize}

\subsection{Two charge black hole in various dimension}
Now we turn to cases where two chemical potentials are turned on. Differential equations are\footnote{ These form of equations are different than the form written in\cite{Jain:2009uj} , because as mentioned earlier that we have done a rescaling.}
\begin{eqnarray}
&&(\phi_1)^{\prime\prime} + ~\left(\dfrac{f^\prime}{f} + 2 \dfrac{H_1^\prime}{H_1} -  \dfrac{\cal H^\prime}{\cal H} -\frac{c}{u} \right) (\phi_1)^\prime\nonumber \\
 &&- \dfrac{a u^{b}(1+k_1)(1+k_2)}{fH_1^2}
\Big[ k_1 \phi_1 + \sqrt{k_1 k_2}~\phi_2\Big] = 0, \label{general2charge}
\end{eqnarray} 
and
\begin{eqnarray}
&&(\phi_2)^{\prime\prime} + ~\left(\dfrac{f^\prime}{f} + 2 \dfrac{H_2^\prime}{H_2} -  \dfrac{\cal H^\prime}{\cal H} -\frac{c}{u} \right) (\phi_2)^\prime\nonumber \\
 &&- \dfrac{a u^{b}(1+k_1)(1+k_2)}{fH_2^2}
\Big[ k_2 \phi_2 + \sqrt{k_1 k_2}~\phi_1\Big] = 0, 
\end{eqnarray} 
Note that ${\cal K}_{i} = \Big[\frac {1}{8\pi G_{d+1}}\sqrt{\frac{g_{uu}}{-g_{tt}}}N_i(u)\Big]_{u\to 1} $  where $N_i=\frac{f H_i^{2}}{u^{m}\cal{H}}$.
Now we compute case by case.
\begin{itemize}
 \item \textbf{D=4:} Here one has $c=0, a=1,b=2$ and $m=0$. Solutions are 
\begin{equation}
 \phi_1 = \frac{(a_0 + \frac{2 a_0 k_1 - b_0 \sqrt{k_1 k_2}}{3} u)}{1+k_1 u}~,~~~~~~~   \phi_2 = \frac{(b_0 + \frac{2 b_0 k_2 - a_0 \sqrt{k_1 k_2}}{3}u)}{1+k_2 u},
 \end{equation}
and
\begin{equation}
{\cal K}_{i}=\frac{N_c^{\frac{3}{2}}(1+k_i)^{2}}{24\pi\sqrt{2(1+k_1)(1+k_2)}}.
\end{equation}
 Using these we get following form of conductivity.
\[ \left( \begin{array}{cc}
\frac{N_c^{\frac{3}{2}}(9 + (12 + k_2) k_1 + 4 k_1^{2})}{6^3 \pi \sqrt{2 (1 + k_1) (1 + k_2)}} & -\frac{2 N_c^{\frac{3}{2}}\sqrt{k_1 k_2}(3+k_1+k_2)}{6^{3}\pi \sqrt{2 (1 + k_1) (1 + k_2)}} \\
-\frac{2 N_c^{\frac{3}{2}}\sqrt{k_1 k_2}(3+k_1+k_2)}{6^{3}\pi \sqrt{2 (1 + k_1) (1 + k_2)}} & \frac{N_c^{\frac{3}{2}}(9 + (12 + k_1) k_2 + 4 k_2^{2})}{6^3 \pi \sqrt{2 (1 + k_1) (1 + k_2)}} \end{array} \right).\]
 
\end{itemize}

\begin{itemize}
 \item \textbf{D=5:} Here we have $c=0, a=1,b=1$ and $m=0$. In this case solutions are
\begin{equation}
  \phi_1 = \frac{(a_0 +\frac{a_0 k_1 - b_0 \sqrt{k_1 k_2}}{2}u)}{1+k_1 u}~,~~~~~~~~~\phi_2 = \frac{(b_0 + \frac{b_0 k_2 - a_0 \sqrt{k_1 k_2}}{2}u)}{1+k_2 u}.
 \end{equation}
Where as
\begin{equation}
{\cal K}_{i} =\frac{N_c^{2}T_{0}(1+k_i)^{2}}{16 \pi \sqrt{(1 + k_1)(1 + k_2)}}.
\end{equation} 
So we get conductivity as
\[ \left( \begin{array}{cc}
\frac{(4 + k_1^{2}+k_1(4 + k_2))N_c^{2}T_o}{64\pi\sqrt{(1 + k_1)(1 + k_2)}} & -\frac{(4 +k_1+k_2)N_c^{2}T_o}{64\pi} \sqrt{\frac{k1 k2}{(1 + k1)(1+k2)}} \\
-\frac{(4 +k_1+k_2)N_c^{2}T_o}{64\pi} \sqrt\frac{k1 k2}{(1 + k1)(1+k2)} & \frac{(4 + k_2^{2}+k_2(4 + k_1))N_c^{2}T_o}{64\pi\sqrt{(1 + k_1)(1 + k_2)}} \end{array} \right).\] 
So $\sigma $  increases linearly with $ T_{H} $.
\end{itemize}

\begin{itemize}
 \item \textbf{D=7:} In this case $c=-1,a=4,b=3$ and $m=1$. Solutions are
\begin{equation}
  \phi_1 = \frac{(a_0 +\frac{a_0 k_1 -2 b_0 \sqrt{k_1 k_2}}{3}u^2)}{1+k_1 u^2}~,~~~~~~~~~~~\phi_2 = \frac{(b_0 + \frac{b_0 k_2 - 2 a_0 \sqrt{k_1 k_2}}{3}u^2)}{1+k_2 u^2}
 \end{equation}
Now
\begin{equation}
{\cal K}_{i} =\frac{4 N_c^{3}T_{0}^{3}(1+k_i)^{2}}{81\sqrt{(1 + k_1)(1 + k_2)}}.
\end{equation} 
Using these one finds conductivity matrix as
\[ \left( \begin{array}{cc}
\frac{4(9+k_1(k_1+4k_2+6))N_c^{3}T_o^{3}}{3^6\sqrt{(1 + k_1)(1 + k_2)}} & -\frac{(6 +k_1+k_2)N_c^{3}T_o^{3}}{3^{6}} \sqrt{\frac{k1 k2}{(1 + k1)(1+k2)}} \\
 -\frac{(6 +k_1+k_2)N_c^{3}T_o^{3}}{3^{6}} \sqrt{\frac{k1 k2}{(1 + k1)(1+k2)}} & \frac{4(9+k_2(k_2+4k_1+6))N_c^{3}T_o^{3}}{3^6\sqrt{(1 + k_1)(1 + k_2)}} \end{array} \right).\]
\end{itemize}
\begin{itemize}
 \item Notice that off diagonal components of the conductivity matrix are negative but they are important, and plays crucial role when we construct thermal conductivity to viscosity ratio.
\end{itemize}
\begin{itemize}
 \item Observe that off diagonal components goes as $\sigma_{ij} \sim \Omega_{i} \Omega_{j} T^{d-3}$ , where $\Omega_{i}= \frac{\mu_{i}}{2 \pi T} $. So switching off one of the chemical potential will make it zero, where as diagonal parts of conductivity goes as $\sigma_{ii} \sim T^{d-3}f_{ii}(\Omega_{1},\Omega_{2}) $, where $ f_{ii}(0,0)\neq 0$ ($\mu =0 $, implies total charge density is zero i.e. there exist equal number of positive as well as negative charge and applying external electric field will induce flow of both in opposite direction which will contribute to electrical current). Since charged particles moves in opposite direction, there will be collisions among them and it ensures finite conductivity. As one increases $\mu$, conductivity should increase as relative number of collisions between opposite charges are less compared to zero chemical potential case. 
 \end{itemize}

\subsection{Three charge black hole in various dimension}
Now we turn to three charge black hole cases.
General form of differential equations are 
\begin{equation}
(\phi_i)^{\prime\prime} + ~\left(\dfrac{f^\prime}{f} + 2 \dfrac{H_i^\prime}{H_i} -  \dfrac{\cal H^\prime}{\cal H} \right) (\phi_i)^\prime- \dfrac{ u^{b}\prod\limits_{j=1}^3 (1+k_j) \sqrt{k_i}}{fH_i^2}
\Big[ \sum\limits_{j=1}^3 \sqrt{k_j}\phi_j \Big] = 0. \label{general3charge}
\end{equation} 
\begin{itemize}
 \item \textbf{D=4:} For this case the one gets $ b=2$ . Relevant results in this case are
 \begin{equation}
  \phi_i = \frac{\Big[3 \phi_i^{0} +\sqrt{k_i} \Big(3 \sqrt{k_i}\phi_i^{0} - \sum\limits_{j=1}^3 \sqrt{k_j}\phi_j^{0}\Big) u\Big]}{3 (1+k_i u)}~,~~{\cal K}_{i}=\frac{N_c^{\frac{3}{2}}(1+k_i)^{2}}{24\pi\sqrt{2\prod\limits_{j=1}^3 (1+k_j)}},
 \end{equation} where $\phi_i^{0}$ is the boundary value of perturbed gauge field.

Let us introduce, $\sigma_{ij} = \frac{N_c^{\frac{3}{2}}}{6^3 \pi \sqrt{2 (1 + k_1) (1 + k_2)(1+k_3)}} \sigma_{ij}^{0}$, where $\sigma_{ij}^{0}$, is given by
\textbf{\begin{scriptsize}\[ \left( \begin{array}{ccc}
9 + (12 + k_2+k_3) k_1 + 4 k_1^{2} & -\sqrt{k_1 k_2}(6+2 k_1+2 k_2-k_3) & -\sqrt{k_1 k_3}(6+2 k_1+2 k_3-k_2)\\
\\
-\sqrt{k_1 k_2}(6+2 k_1+2 k_2-k_3) & 9 + (12 + k_1+k_3) k_2 + 4 k_2^{2} &-\sqrt{k_2 k_3}(6+2 k_2+2 k_3-k_1)\\
\\
-\sqrt{k_1 k_3}(6+2 k_1+2 k_3-k_2)& -\sqrt{k_2 k_3}(6+2 k_2+2 k_3-k_1) & 9 + (12 + k_1+k_2) k_3 + 4 k_3^{2}\end{array}\right).\]                                                                                                                                  \end{scriptsize}}
\end{itemize}

\begin{itemize}
 \item \textbf{D=5:} For this case $b=1$. Results needed for conductivity calculation are
\begin{equation}
  \phi_i = \frac{\Big[2 \phi_i^{0} +\sqrt{k_i} \Big(2 \sqrt{k_i}\phi_i^{0} - \sum\limits_{j=1}^3 \sqrt{k_j}\phi_j^{0}\Big) u\Big]}{2 (1+k_i u)},~~~~~
 {\cal K}_{i} =\frac{N_c^{2}T_{0}(1+k_i)^{2}}{16 \pi \sqrt{(1 + k_1)(1 + k_2)(1+k_3)}}.
 \end{equation}

Defining as before $\sigma_{ij} = \frac{N_c^{2}T_0}{64 \pi \sqrt{ (1 + k_1) (1 + k_2)(1+k_3)}} \sigma_{ij}^{0}$, where $\sigma_{ij}^{0}$ is given by
\\

\textbf{\begin{scriptsize}\[ \left( \begin{array}{ccc}
4 + k_1^{2}+k_1(4 + k_2+k_3) & -\sqrt{k_1 k_2}(4 +k_1+k_2-k_3) & -\sqrt{k_1 k_3}(4 +k_1+k_3-k_2) \\
\\
 -\sqrt{k_1 k_2}(4 +k_1+k_2-k_3)& 4 + k_2^{2}+k_2(4 + k_1+k_3)&-\sqrt{k_2 k_3}(4 +k_2+k_3-k_1)  \\
\\
-\sqrt{k_1 k_3}(4 +k_1+k_3-k_2)&-\sqrt{k_2 k_3}(4 +k_2+k_3-k_1)& 4 + k_3^{2}+k_3(4 + k_1+k_2)\end{array} \right)\]                                                                                                                     \end{scriptsize}
}
\end{itemize}

\subsection{Four charge black hole}
Differential equations are 
\begin{equation}
(\phi_i)^{\prime\prime} + ~\left(\dfrac{f^\prime}{f} + 2 \dfrac{H_i^\prime}{H_i} -  \dfrac{\cal H^\prime}{\cal H} \right) (\phi_i)^\prime- \dfrac{ u^{2}\prod\limits_{j=1}^4 (1+k_j) \sqrt{k_i}}{fH_i^2}
\Big[ \sum\limits_{j=1}^4 \sqrt{k_j}\phi_j \Big] = 0,
\end{equation} and solutions are

 \begin{equation}
  \phi_i = \frac{\Big[3 \phi_i^{0} +\sqrt{k_i} \Big(3 \sqrt{k_i}\phi_i^{0} - \sum\limits_{j=1}^4 \sqrt{k_j}\phi_j^{0}\Big) u\Big]}{3 (1+k_i u)}~,~~{\cal K}_{i}=\frac{N_c^{\frac{3}{2}}(1+k_i)^{2}}{24\pi\sqrt{2\prod\limits_{j=1}^4 (1+k_j)}}.
 \end{equation} 

Using these we get following form of conductivity.

\begin{equation}
\sigma_{ij}= \frac{N_c^{\frac{3}{2}}}{6^3 \pi \sqrt{2 (1 + k_1) (1 + k_2)(1+k_3)(1+k_4)}}\sigma^{0}_{ij}.
\end{equation}
Where
\begin{equation}
\sigma^{0}_{ii}=9+\Big(12+\sum\limits_{j=1}^4 k_j\Big)k_i+3 k_i^{2}~~~~\nonumber\\\rm{and}~~
\sigma^{0}_{ij}=-\Big(6-\sum\limits_{l=1}^4 k_l +3\Big(k_i +k_j\Big)\Big).\nonumber\\
\end{equation}

\begin{itemize}
 \item \textbf{Some special cases :} Using above results one can study  special cases such as effect of small chemical potential or the case with equal chemical potential. Note that in the case when all the chemical potential are equal then there exist no second order phase transition. In this case, temperature i.e.  $T \geq 0$ gives a constraint on the possible maximum value of chemical potential. 

\begin{center}
\begin{tabular}{l*{2}{c}r}
 Dimension         & Constraint($T \geq 0$)    &$\sigma $ \\
\hline
\\
  5           &$ k\leq 2 $ & $ \frac{3 (2 - k)^2  N_{c}^2 T_0}{32 \pi (1 + k)^\frac{3}{2} }$ \\
\\ 
 4          &$ k\leq 3$& $ \frac{(3 - k)^2 N_{c}^\frac{3}{2}}{27 \sqrt{2}(1 + k)^2 \pi}$ \\
\\ 
 7          &$k\leq 3 $& $\frac{16 (3 - k)^2 N_{c}^3 T_{0}^3}{729 (1 + k)}$ \\
\end{tabular}             \end{center}
Note that as $T \to 0 $, $\sigma\to 0 $ quadratically in the parameter $ k $, irrespective of which dimension we are in\footnote{ Determinant of conductivity matrix for general $\mu$ also vanishes in similar way once we approach extremality even for M2 brane case where conductivity is independent of temperature.}.  
\end{itemize}

\section{General analysis for electrical conductivity}
In the absence of chemical potential the general form of conductivity was written down in \cite{Iqbal:2008by} using membrane paradigm at the horizon (see also \cite{Kovtun:2003wp},\cite{Kovtun:2008kx} for other general discussions). But in the presence of chemical potential one can't use those formulas as there is a nontrivial flow from horizon to boundary. To obtain a general form  of conductivity, we analyze the most general form of the perturbation equation both near the horizon as well as near the boundary and use these information's to fix the form of conductivity. Let us focus at the single charge case first.
\begin{itemize}
 \item \textbf{Single charge :} In this case equation is
\begin{equation}
 \frac{d^2}{du^2}\phi^i(u)+(\frac{d}{du}\log(\sqrt{-g}G_{ii}g^{xx}g^{uu}))\frac{d}{du}\phi^i(u)+\frac{1}{g_{tt}}G_{ii}(F_{ut}^i)^{2}\phi^i(u)=0,\label{generalwithchemical1}
\end{equation}
Near the horizon ($u\rightarrow1$), $N_{i}=\sqrt{-g}G_{ii}g^{xx}g^{uu}\rightarrow (1-u)A$, and $\frac{1}{g_{tt}}G_{ii}(F_{ut}^i)^{2}\rightarrow -\frac{F}{1-u}$ where $A, F$ are some constant\footnote{Lets take $g_{tt}=(1-u) \gamma_{0} $, then $F=(\frac{1}{\gamma_{0}}G_{ii}(F_{ut}^i)^{2})|_{u=1}$.}.
To the leading order in inverse power of $(1-u)$, one gets
\begin{equation}
 \frac{d}{du}\phi^i+F\phi^i=0.
\end{equation}
Using this the general form of solution can be given by
\begin{equation}
 \phi^i(u)=a_{0}\frac{n H_{i}+(H_{i}F-H_{i}^{'})(1-u^{n})}{H_{i}(u)(H_{i}F+n H_{i}-H_{i}^{'})}
\end{equation}
Where $H_{i},H_{i}^{'}$ are evaluated at the horizon\footnote{Note that $H_{i}(u)$ can be replaced by 
\begin{equation}
\frac{\sqrt{\sqrt{g}G_{ii}g^{uu}g^{tt}u^{n-1}}}{\sqrt{\sqrt{g}G_{ii}g^{uu}g^{tt}u^{n-1}|_{u=0}}}.
\end{equation}, which makes expression for conductivity independent of background.}($u=1$).
The general form of $\sigma$ for single charge case can be written as
\begin{equation}
\sigma =\frac{1}{8\pi G_{d+1}}G_{II}g^{xx}\sqrt{\frac{-g}{g_{uu}g_{tt}}}(\frac{n}{H_{i}F+n H_{i}-H_{i}^{'}})^{2},
\end{equation}
where every quantity in the above is calculated at the horizon\footnote{In all the cases considered above $n H_{i}-H_{i}^{'}=n$. In the absence of chemical potential $F=0$ so one obtains the usual result as written in \cite{Iqbal:2008by}, see also \cite{Myers:2009ij} for related discussion in the presence of chemical potential.}($u=1$). Value of $n$ can be determined by looking at 
\begin{equation}
\frac{d}{du}(N_i\frac{d}{du}\phi^i(u))+\sum\limits_{j=1}^m  M_{ij}\phi^j(u)=0,
\end{equation}  
near the boundary i.e. $u\sim0$, if
\begin{equation}
N_{i}\sim u^{1-n}
\end{equation}
then
\begin{equation}
\phi^i(u)\sim a_0 + a_1 u^{n}.
\end{equation} 
As an example consider single charge $M5-brane $ case. Here
\begin{equation}
N_{1}\sim u^{-1}\Rightarrow n=2\nonumber\\ ~~~and~~~~
F=\dfrac{ 4 k_1 (1+k_1)}{(3+k_1)H_1^2}
\end{equation}
which after plugging on the expression for conductivity one finds same result as in section $4.1.$ Note that explicit solution in the case of $M5-brane$
has $\phi(u)=\phi_{0}\frac{1+\frac{ k u^{2}}{3}}{1+k u^{2}}$ i.e. $ n = 2$, which is consistent with our analysis.
\end{itemize}

\begin{itemize}
 \item \textbf{Multicharge case:} Now we analyze multi charge black hole (one can do an analogous analysis as done above for multicharge case, but here we avoid repetition). The conductivity matrix that we obtained is symmetric, hence one can go to a suitable basis where conductivity is diagonal. In this basis there is no effective interaction between different fields, in other words they behave as independent field. Which implies we can apply same  general formula for conductivity as for single charge case. As an example lets take $D3$ brane $2$ charge case. Here the diagonal form of the conductivity is given by
\[ \left( \begin{array}{cc}
 \frac{N_c^{2}T_o}{16\pi\sqrt{(1 + k_1)(1 + k_2)}} & 0 \\
 0  & \frac{(2+k_1+k_2)^{2}N_c^{2}T_o}{64\pi\sqrt{(1 + k_1)(1 + k_2)}} \end{array} \right)\] 
   Expression for  charge density and chemical potential in this basis are given by
\begin{equation}
\varrho_1 =\frac{\sqrt{k_2} \rho_1}{\sqrt{k_1 + k_2}} + \frac{\sqrt{k_1} \rho_2}{\sqrt{k_1 + k_2}},~~ \nonumber\\
  \varrho_2 =-\frac{\sqrt{k_1} \rho_1}{\sqrt{k_1 + k_2}} + 
  \frac{\sqrt{k_2} \rho_2}{\sqrt{k_1 + k_2}} 
\end{equation}

\begin{equation}
\tilde{\mu_1}  =\frac{\sqrt{k_2} \mu_1}{\sqrt{k_1 + k_2}} + \frac{\sqrt{k_1} \mu_2}{\sqrt{k_1 + k_2}} ,~~~
\tilde{\mu_2}=-\frac{\sqrt{k_1} \mu_1}{\sqrt{k_1 + k_2}} + \frac{\sqrt{k_2} \mu_2}{\sqrt{k_1 + k_2}} ~.
\end{equation}
One can start with differential equation and do a same rotation as above to go to a diagonal basis and using general results of single charge, can obtain same values of conductivities as appeared in the diagonal elements.
\end{itemize}

\section{Thermal conductivity}
In this section we first review hydrodynamics with multiple conserved charges and write down expression for thermal conductivity, using which we compute thermal conductivity and  show that thermal conductivity and viscosity obeys Wiedemann-Franz like law even in the presence of multiple chemical potential. Single charge case was discussed in  \cite{Son:2006em} .
\begin{itemize}
 \item \textbf{\textit{Relativistic hydrodynamics with multiple conserved charge:}}
 The continuity equations are
\begin{equation}
   \d_\mu T^{\mu\nu}=0, ~~~ \d_\mu J_{i}^\mu =0
\end{equation}
where 
\begin{equation}
   T^{\mu\nu} = (\epsilon+P) u^\mu u^\nu + P g^{\mu\nu}+\tau^{\mu\nu}
    ,~~~
    J_{i}^\mu = \rho_{i} u^\mu + \nu_{i}^\mu\label{velo}
\end{equation}
In the above $\epsilon$ and P are the local energy density and pressure, $u^\mu$ is the local velocity and it obeys $ u_\mu u^\mu = -1$, where as $\tau^{\mu\nu}$ and $\nu_{i}^\mu$ are the dissipative parts of stress-energy tensor and current.

 Following \cite{Son:2006em} one can choose
 $u^\mu$ and $\rho_{i}$'s so that 
\begin{equation}\label{transverse}
    u_\mu \tau^{\mu\nu} = u_\mu \nu^\mu_{i} = 0\,.
\end{equation}
Note that
\begin{equation}\label{udT}
    u_\nu \d_\mu T^{\mu\nu} = - (\epsilon+P)\d_\mu u^\mu
    - u^\mu \d_\mu \epsilon + u_\nu \d_\mu \tau^{\mu\nu} =0
\end{equation}
We also have 
\begin{equation}
    \epsilon+P = Ts +\sum\limits_{i=1}^m \mu^{i} \rho_{i}, \qquad d\epsilon = Tds +\sum\limits_{i=1}^m \mu^{i} d\rho_{i}\label{thermo}
\end{equation}
Using which one gets
\begin{equation}
    \d_\mu \left(su^\mu - \sum\limits_{i=1}^m\frac{\mu^{i}}{ T}\nu^\mu_{i}\right) =
    -\sum\limits_{i=1}^m\nu^\mu_{i} \d_\mu \frac{\mu^{i}}{T} - \frac{\tau^{\mu\nu}}{T} \d_\mu u^\mu\,.
\end{equation}
Now $\d_\mu \left(su^\mu - \sum\limits_{i=1}^m\frac{\mu^{i}}{ T}\nu^\mu_{i}\right)$ can be interpreted as the divergence of the entropy current which implies right hand side is positive. So we write
\begin{equation}
 \nu^{\mu}_i = -\sum\limits_{j=1}^m\varkappa_{ij} \left(\d^\mu\frac{\mu^{j}}{ T}+ u^\mu u^\lambda \d_\lambda \frac{\mu^{j}}{ T}\right)
\end{equation} and similarly for $\tau^{\mu\nu}$ (see \cite{Son:2006em}) .
 To interpret $ \varkappa_{ij} $  as the
 coefficient of thermal conductivity, consider no charge current i.e. $ J^\alpha_{j} = 0, $\footnote{In our notation $\mu ,\nu$ runs from $t,1,2...D$ , where as $\alpha$ runs from $1,2,...D$, and i,j are R-charge indices.}  but there is an energy flow,
  $ T^{t\alpha}\neq 0 $, which is the heat flow.  Take $ u^\alpha $ to be small so that one gets using eqn. (\ref{velo}) 
\begin{equation}
 \rho_{i} u^\alpha=\sum\limits_{j=1}^m\varkappa_{ij} \d^\alpha\frac{\mu^{j}}{T}.\nonumber\\
\end{equation}

From which one can obtain
\begin{equation}
 \sum\limits_{i,j=1}^m\rho_{i}\varkappa_{ij}^{-1} \rho_{j} u^\alpha = \sum\limits_{i=1}^m\rho_{i}\d^\alpha \frac{\mu^{i}}{T}\,,
\end{equation}
hence 
\begin{equation}
  u^\alpha =\frac{1}{\sum\limits_{i,j=1}^m\rho_{i}\varkappa_{ij}^{-1}\rho_{j}} \sum\limits_{l=1}^m\rho_{l}\d^\alpha \frac{\mu^{l}}{T}\,.
\end{equation}
 Using eqn. (\ref{thermo}) we get
\begin{equation}
\sum\limits_{i=1}^m\rho_{i}\d^\alpha \frac{\mu^{i}}{T}=-\frac{\epsilon+P}{T^{2}}\d^\alpha T+\frac{\d^\alpha P}{T}
\end{equation}
after substitution, this gives
\begin{equation}
  u^\alpha =-\frac{1}{\sum\limits_{i,j=1}^m\rho_{i}\varkappa_{ij}^{-1}\rho_{j}}(\frac{\epsilon+P}{T^{2}})(\d^\alpha T-\frac{T}{\epsilon+P}\d^\alpha P) \,.
\end{equation}
 Therefore
  \begin{equation}
     T^{t\alpha} = (\epsilon+P) u^\alpha = -\frac{1}{\sum\limits_{i,j=1}^m\rho_{i}\varkappa_{ij}^{-1}\rho_{j}}(\frac{\epsilon+P}{T})^{2}(\d^\alpha T-\frac{T}{\epsilon+P}\d^\alpha P)
 \end{equation}
hence the coefficient of thermal conductivity can be read off as
 \begin{equation}
   \kappa_T =\left( \frac{\epsilon+P}{ T}\right)^2\frac{1}{\sum\limits_{i,j=1}^m\rho_{i}\varkappa_{ij}^{-1}\rho_{j}}\label{thermalconductivity}.
 \end{equation}
Note that, $\varkappa_{ij}$ can be found out from greens function as 
\begin{equation}
   G_{ij} (\omega, q=0) = -i \omega \frac{\varkappa_{ij}}{ T}\,,
 \end{equation}
where $ J^{\alpha}_{i}=-G_{ij}(\omega,q=0)A^{\alpha}_j $  and $ G_{ij}(\omega,q=0)$ can be obtained using kubo formula as done in previous sections.
Now we turn to computation of $\kappa_T$, and dimension less ratio \footnote{There exist interesting relationship 
\begin{equation}
\frac{\sum\limits_{j=1}^m(\mu^{j})^{2}}{\sum\limits_{i,j=1}^m\rho_{i}\varkappa_{ij}^{-1}\rho_{j}}=\frac{\sum\limits_{i,j=1}^m\mu_{i}\varkappa_{ij}\mu_{j}}{\sum\limits_{j=1}^m(\rho_{j})^{2}}\nonumber\\ 
\end{equation}
in all the example considered in the text}
\begin{equation}
\frac{\kappa_T}{\eta T}\sum\limits_{j=1}^m(\mu^{j})^{2} 
\end{equation}
 case by case.
\end{itemize}

\begin{itemize}
 \item \textbf{\textit{Single charge black hole:}}
Note that for single charge black hole $\frac{1}{\rho_{i}\varkappa_{ij}^{-1}\rho_{j}}=\frac{\varkappa}{\rho^{2}}$. So that one gets
\begin{equation}
   \kappa_T =\left( \frac{\epsilon+P}{ \rho T}\right)^2\varkappa=\left( \frac{\epsilon+P}{ \rho }\right)^2\frac{\sigma}{T}\label{themalcon}
 \end{equation}
Rather than providing details we here tabulate the thermal conductivity and dimension less ratio $\frac{\kappa_T\mu^{2}}{\eta T}$, where $\eta$
is the shear viscosity.\\
 \begin{center}
\begin{tabular}{l*{2}{c}r}
 Dimension         &  $\kappa_T$   &$\frac{\kappa_T\mu^{2}}{\eta T}$ \\
\hline
  5           &$\frac{(1 + k)^{2} N_c^2 T^2 \pi}{ k (2 + k)}$&$8 \pi^{2}$ \\
  4          &$\frac{2 \sqrt{2} (1 + k)^{3/2} N_c^{3/2} T \pi}{3 k}$&$32 \pi^{2}$ \\
  7          &$\frac{8 (1 + k)^{3} N_c^3 T^{4} \pi^{2}}{3 k (3 + k)^{3}}$&$2 \pi^{2}$ \\
\end{tabular}             \end{center}
\end{itemize}

\begin{itemize}
\item \textbf{\textit{Two charge black hole:}}
In the ratio $\frac{\kappa_T\mu^{2}}{\eta T}$, $\mu^{2}$ is replaced by $\mu_{1}^{2}+\mu_{2}^{2}$. Note that $ \mu_i\to -\mu_i$ is a symmetry \footnote{ The expression for $\kappa_T$ in eqn. (\ref{thermalconductivity}) is invariant under $SO(m)$ rotation among $\rho_{i}$'s.} which implies reversing the sign of charge density.  
 \begin{center}
\begin{tabular}{l*{2}{c}r}
 Dimension         &  \textbf{$\kappa_T$}   &\textbf{$\frac{\kappa_T(\mu_{1}^{2}+\mu_{2}^{2})}{\eta T}$} \\
\hline
  &~~~~~~~~~~~~~~~~~~~~~~~~~~~~~~~~~~~~~~~~~~~~~~~~~~~~~~~~~~~~~~~~~~~\\ 
  5           &$\frac{N_c^{2} T^{2} \pi }{ \Big(2 +\sum\limits_{j=1}^2 k_j  \Big)\Big(\sum\limits_{j=1}^2 \frac{ k_j}{(1+k_j)^{2}}\Big)} $&$8 \pi^{2}$ \\
  & ~~~~~~~~~~~~~~~~~~~~~~~~~~~~~~~~~~~~~~~~~~~~~~~~~~~~~~~~~~~~~~~~~~~\\
  4          &$\frac{(2 N_c)^{\frac{3}{2}} T \pi}{\Big(\sum\limits_{j=1}^2 \frac{3 k_j}{(1+k_j)^{2}}\Big)\sqrt{\prod\limits_{i=1}^2 (1+k_i)}} $ &$32 \pi^{2}$ \\
  & ~~~~~~~~~~~~~~~~~~~~~~~~~~~~~~~~~~~~~~~~~~~~~~~~~~~~~~~~~~~~~~~~~~~~\\
  7          &$\frac{(2 N_c)^{3} T^{4} \pi^{2}\prod\limits_{i=1}^2 (1+k_i)}{\Big(\sum\limits_{j=1}^2 \frac{3 k_j}{(1+k_j)^{2}}\Big)\Big(3 +\sum\limits_{j=1}^2 k_j  -\prod\limits_{j=1}^2 k_j \Big)^{3}} $   &$2 \pi^{2}$ \\
\end{tabular}             \end{center}
\end{itemize}
\begin{itemize}
\item \textbf{\textit{Three charge black hole:}}

\begin{center}
\begin{tabular}{l*{2}{c}r}
Dimension &  \textbf{$\kappa_T$}   &\textbf{$\frac{\kappa_T(\mu_{1}^{2}+\mu_{2}^{2}+\mu_{3}^{2})}{\eta T}$} \\
\hline
  &~~~~~~~~~~~~~~~~~~~~~~~~~~~~~~~~~~~~~~~~~~~~~~~~~~~~~~~~~~~~~~~~~~~\\ 
  5&$\frac{ N_c^{2} T^{2} \pi}{\Big(2 +\sum\limits_{j=1}^3 k_j  -\prod\limits_{j=1}^3 k_j\Big)\Big(\sum\limits_{j=1}^3 \frac{ k_j}{(1+k_j)^{2}}\Big) }$&$8 \pi^{2}$ \\
  & ~~~~~~~~~~~~~~~~~~~~~~~~~~~~~~~~~~~~~~~~~~~~~~~~~~~~~~~~~~~~~~~~~~~\\
  4       &$\frac{(2 N_c)^{\frac{3}{2}} T \pi}{\Big(\sum\limits_{j=1}^3 \frac{3 k_j}{(1+k_j)^{2}}\Big)\sqrt{\prod\limits_{i=1}^3 (1+k_i)}} $ &$32 \pi^{2}$ \\
  
\end{tabular}             \end{center}	
\end{itemize}
\begin{itemize}
\item \textbf{\textit{Four charge black hole (4 Dimensional black hole):}}
The thermal conductivity is given by

\begin{equation}
  \kappa_T =\frac{(2 N_c)^{\frac{3}{2}} T \pi}{\Big(\sum\limits_{j=1}^4 \frac{3 k_j}{(1+k_j)^{2}}\Big)\sqrt{\prod\limits_{i=1}^4 (1+k_i)}}
 \end{equation}
and
 \begin{equation}
\frac{\kappa_T(\mu_{1}^{2}+\mu_{2}^{2}+\mu_{3}^{2}+\mu_{4}^{2})}{\eta T}=32 \pi^{2}.
\end{equation}
\end{itemize}

\begin{itemize}
 \item \textbf{\underline{$\mu = 0$~}:} We observe that irrespective of number of chemical potential turned on, thermal conductivity to viscosity ratio shows same value. We also observe that as $\mu\to 0$ i.e. $\rho\to 0$, thermal conductivity given in eqn. (\ref{themalcon}), diverges which implies finite, non decaying momentum (see \cite{Vojta}, for more details). 

\end{itemize}

\section{Conductivity for extremal black hole}
In this section we compute electrical conductivity for extremal background. At extremality
metric in the vicinity of horizon takes the form
\begin{equation}
\bar g_{tt} = - f(u) A_1(u) = - (1-u)^{2} \gamma_{0}, ~~\bar g_{uu} = A_2(u) f^{-1}(u) = \frac{\gamma_{u}}{(1-u)^{2}} ,~f(u)=(1-u)^2 V(u).
\end{equation}
where $\gamma_{0}$ and $\gamma_{u}$ are some constants.
Near the horizon eqn. (\ref{eqnmotion}) reduces to
\begin{equation}
\frac{d^2}{du^2}\phi_i(u)-\frac{2}{1-u}\frac{d}{du}\phi_i(u)+\frac{\gamma_{u}}{\gamma_{0}} \frac{\omega^{2}}{(1-u)^{4}}\phi_i(u)-\frac{c_i}{(1-u)^{2}}\frac{(\sum\limits_{j=1}^m d_j \phi_j(u))}{\gamma_{0}} = 0\label{nearhorizon}
\end{equation}
Note that $c_i=F_{ut}^{i}(u=1)$ and $ d_j= G_{jj}(u)F_{ut}^j(u)$ at $ u=1 $. Following \cite{Faulkner:2009wj},\cite{Edalati:2009bi}  let us define $u=1-\frac{\omega}{\xi}$. In this coordinate system eqn. (\ref{nearhorizon}) reduces to
\begin{equation}
\frac{d^2}{d\xi^2}\phi_i(\xi)+\frac{\gamma_{u}}{\gamma_{0}}\phi_i(\xi)-\frac{c_i}{\xi^2}\frac{(\sum\limits_{j=1}^m d_j \phi_j(\xi))}{\gamma_{0}}=0\label{nearhorizon1}
\end{equation}
Above equation is in general a complicated coupled differential equation. To solve it we observe that
\begin{equation}
\frac{\frac{d^2}{d\xi^2}\phi_i(\xi)+\frac{\gamma_{u}}{\gamma_{0}}\phi_i(\xi)}{c_i}=\frac{(\sum\limits_{j=1}^m d_j \phi_j(\xi))}{\gamma_{0} \xi^{2}}\label{nearhorizon2}
\end{equation}
In the case when more than one field is present then we get
\begin{equation}
\frac{\frac{d^2}{d\xi^2}\phi_1(\xi)+\frac{\gamma_{u}}{\gamma_{0}}\phi_1(\xi)}{c_1}=\frac{\frac{d^2}{d\xi^2}\phi_2(\xi)+\frac{\gamma_{u}}{\gamma_{0}}\phi_2(\xi)}{c_2}=....\label{nearhorizon3}
\end{equation}
We take solution of the form 
\begin{equation}
\frac{\phi_1(\xi)}{c_1}=\frac{\phi_2(\xi)}{c_2}=....\label{nearhorizon4}
\end{equation}
Plugging eqn. (\ref{nearhorizon4}) in eqn. (\ref{nearhorizon1}) one obtains
\begin{equation}
\frac{d^2}{d\xi^2}\phi_i(\xi)+\frac{\gamma_{u}}{\gamma_{0}}\phi_i(\xi)-\frac{(\sum\limits_{j=1}^m d_j c_j)}{\gamma_{0}\xi^2}\phi_i(\xi)=0\label{nearhorizon5}
\end{equation}

Introduce $\eta=\sqrt{\frac{\gamma_{u}}{\gamma_{0}}}~ \xi $ and $ a=\frac{(\sum\limits_{j=1}^m d_j c_j)}{\gamma_{0}} $, so that one gets ( from eqn.(\ref{nearhorizon5}))
\begin{equation}
\frac{d^2}{d\eta^2}\phi_i(\eta)+\phi_i(\eta)-\frac{a}{\eta^2}\phi_i(\eta)=0\label{nearhorizon6}
\end{equation}
The incoming solution to eqn. (\ref{nearhorizon6}) takes the form
\begin{equation}
\phi_i(\eta)= C H^{1}_{\nu}(\eta),
\end{equation}
where  $H^{1}_{\nu}(\eta)$ is henkel function and $\nu=\frac{\sqrt{1+4 a}}{2}$. 
Taking $\eta\rightarrow 0$ limit one gets
\begin{equation}
\lim_{\eta\rightarrow 0}\phi_i(\eta)=\eta^{\frac{1}{2}+\nu}2^{-\nu}(\frac{1}{\Gamma[1+\nu]}-i \frac{\cos(\pi\nu)\Gamma[-\nu]}{\pi})-i\eta^{\frac{1}{2}-\nu}2^{\nu}\frac{\Gamma[\nu]}{\pi}
\end{equation}

Using $\eta =\sqrt{\frac{\gamma_{u}}{\gamma_{0}}}\frac{\omega}{(1-u)}, $ and some properties of  Gamma functions as well as doing some rescaling one finds
\begin{equation}
\phi_i(u\rightarrow 1)=A_0\Big[\frac{1}{(1-u)^{\frac{1}{2}-\nu}}+(\sqrt{\frac{\gamma_{u}}{\gamma_{0}}})^{2\nu}(\frac{\omega}{2})^{2\nu}\frac{\pi(i-\cot(\nu\pi))}{\Gamma[1+\nu]\Gamma[\nu]}\frac{1}{(1-u)^{\frac{1}{2}+\nu}}\Big].
\end{equation}
Again using properties of Gamma functions we get
\begin{equation}
\phi_i(u\rightarrow 1)=A_0\Big[\frac{1}{(1-u)^{\frac{1}{2}-\nu}}-(\sqrt{\frac{\gamma_{u}}{\gamma_{0}}})^{2\nu}(\frac{\omega}{2})^{2\nu}\frac{\Gamma[1-\nu]}{\Gamma[1+\nu]}\frac{e^{-i\nu\pi}}{(1-u)^{\frac{1}{2}+\nu}}\Big].
\end{equation}
Upon comparing with standard form \cite{Edalati:2009bi} we conclude
\begin{equation}
g(\omega)=-2\nu e^{-i\nu\pi}(\sqrt{\frac{\gamma_{u}}{\gamma_{0}}})^{2\nu}(\frac{\omega}{2})^{2\nu}\frac{\Gamma[1-\nu]}{\Gamma[1+\nu]}\label{gomega}
\end{equation} 
which corresponds to greens function of operator of dimension $\delta=\nu+\frac{1}{2}$.
We obtain conductivity to be proportional to 
\begin{equation}
\sigma\propto\lim_{\omega\rightarrow 0}\frac{1}{\omega}\Im [g(\omega)]\propto(\omega)^{2\nu -1},
\end{equation}
where 
\begin{eqnarray}
2\nu&& =\sqrt{1+4 a}\nonumber\\
&&=\sqrt{1+ (\frac{4}{\gamma_{0}})\sum\limits_{j=1}^m d_j c_j}\nonumber\\
&&=\sqrt{1+(\frac{4}{\gamma_{0}})\sum\limits_{j=1}^m G_{jj}(F_{ut}^{j})^{2}}\label{nuvalue}.
\end{eqnarray}
In the above expression every quantity is calculated at the horizon ($u=1$).
Hence, we see only way to get non-zero conductivity in the limit $\omega\rightarrow 0 $ at extremality  is $\nu\leq \frac{1}{2}$ where as  $\sigma\rightarrow  0 $ if $ \nu > \frac{1}{2}$.

\begin{itemize}
 \item  To obtain above form of $g(\omega)$, we have only assumed that extremal black hole exhibits double pole. So the expression for operator dimension in general follows only from criteria of extremality i.e. it is independent of particular background. In all the examples considered below
we find $\nu= \frac{3}{2} \Rightarrow \delta = \nu +\frac{1}{2}=2$. There are other classes of black hole as well (dialatonic black hole \cite{Horowitz:2009ij}) where one finds $ \delta=2 $.
\end{itemize}

\begin{itemize}
 \item \textbf{\textit{Four dimensional black hole:}}
In this case
\begin{equation}
 2\nu=\sqrt{1+ 4\frac{\prod\limits_{i=1}^4(1+k_i)}{3  +\sum\limits_{j=1}^4 k_j + \prod\limits_{i=1}^4 k_i  }\Big(\sum\limits_{j=1}^4\frac{k_i}{(1+k_i)^{2}}\Big)}\label{nu4}
\end{equation}
Using extremality condition\footnote{$k_1=\frac{3+2(k_2+k_3+k_4)+k_2(k_3+k_4)+k_3 k_4}{k_2k_3k_3-2-k_2-k_3-k_4}$}(see appendix) we get
$2\nu=3$.
\end{itemize}

\begin{itemize}
 \item \textbf{\textit{Five dimensional black hole:}}
In this case  
\begin{equation}
 2\nu=\sqrt{1+ 4\frac{\prod\limits_{i=1}^3(1+k_i)}{1+\prod\limits_{i=1}^3 k_i }\Big(\sum\limits_{j=1}^3\frac{k_i}{(1+k_i)^{2}}\Big)}\label{nu5}
\end{equation}
Using extremality condition\footnote{$k_3=\frac{2+k_1+k_2}{k_1 k_2-1}$} one finds
$ 2\nu= 3 $. Which implies $\delta=\nu+\frac{1}{2}=2$. Note that above result also applicable for $5$d Rn black hole (for which $k_1=k_2=k_3$) considered in other places \cite{Faulkner:2009wj}.
\end{itemize}

\begin{itemize}
 \item \textbf{\textit{Seven dimensional black hole:}}
In this case
\begin{equation}
 2\nu=\sqrt{1+ 4\frac{4(1+k_1)(1+k_2)}{3+k_1 k_2}(\frac{k_1}{(1+k_1)^{2}}+\frac{k_2}{(1+k_2)^{2}})}\label{nu7}
\end{equation}
Now extremality condition implies $k_1=\frac{3+k_2}{k_2-1}$. So one gets $ 2\nu= 3 $. 
\end{itemize}
\begin{itemize}
 \item Above results implies that for black hole at extremality obeys
\begin{equation}
(\frac{1}{\gamma_{0}})\sum\limits_{j=1}^m G_{jj}(F_{ut}^{j})^{2}=2.
\end{equation}
It would be interesting to find out under what conditions extremal backgrounds obeys this relation.
\end{itemize}

\section{Schr\"{o}edinger like equation with effective potential}
One can gain much clearer understanding by reducing the problem of finding conductivity to solving equivalent Schr\"{o}edinger problem \cite{Horowitz:2009ij}.
General form of perturbation equation in $d+1$ dimensional gravity theory (d-dimensional gauge theory) with multiple chemical potential are
\begin{equation}
\frac{d}{du}(N_i\frac{d}{du}\phi_i(u))-\omega^2 N_i~ g_{uu} g^{tt} \phi_i(u)+\sum\limits_{j=1}^m  M_{ij}\phi_j(u)=0.
\end{equation}
with
\begin{equation}
N_i=\sqrt{-g}G_{ii}g^{xx}g^{uu}. 
\end{equation}
and 
\begin{equation}
 M_{ij}=F_{ut}^i \sqrt{-g}G_{ii}g^{xx}g^{uu}g^{tt}G_{jj}F_{ut}^j.
\end{equation}
Define $f_{i}^{2} = G_{ii}$, $h_{i}^{2} = f_{i}^{2} g_{xx}^{\frac{d-3}{2}} $, new variable $\frac{d}{dz} = \sqrt{-\frac{g_{tt}}{g_{uu}}} \frac{d}{du}$, and new wave function $\Psi_i = h_{i} \phi_{i}   $. Using above definitions we get 
\begin{equation}
 -\Psi_i^{''} + \sum\limits_{j=1}^m  V_{ij}\Psi_j = \omega^2 \Psi_i,\label{schrodinger}
\end{equation} 
 with
\begin{equation}
V_{ij} = \frac{h_{i}^{''}}{h_{i}}\delta_{ij}- f_{i} f_{j} g^{tt} A_{t}^{'i}A_{t}^{'j},
\end{equation}
where $'$ denotes derivative with respect to $ z $ coordinates and $A_{t}^{i}$ 's are background gauge fields. Since $V_{ij}$ is symmetric one can do $So(N)$ rotation among these fields $\Psi$ to go to a diagonal basis where one gets decoupled Schr\"{o}edinger equation. In the remaining part of this section, after commenting about the form of the potential at the horizon and conductivity in terms of reflection coefficient we turn to conductivity for extremal black holes.\begin{itemize}
 \item {\textbf{Potential at the horizon:}} Regularity at the horizon implies
\begin{equation}
\lim_{u\rightarrow 1}\frac{d}{du}\phi_i(u)=-i \omega \lim_{u\to 1}\sqrt{-\frac{g_{uu}}{g_{tt}}} \phi_{i}(u)+{\mathcal{O}}(\omega^{2}). 
\end{equation}
Now using, $\frac{d}{dz} = \sqrt{-\frac{g_{tt}}{g_{uu}}} \frac{d}{du}$ at the horizon we get
\begin{equation}
\frac{d}{dz}\Psi_i(z)=-i \omega \Psi_{i}(z)+{\mathcal{O}}(\omega^{2}). 
\end{equation}
Plugging this in equation (\ref{schrodinger}) we conclude $ V=0$, at the horizon. 
\end{itemize}
Now we can follow standard procedure as in \cite{Horowitz:2009ij} to interpret conductivity in terms of reflection coefficient.\footnote{If we follow effective action approach as developed in section 3, we get
 $\sigma(\omega\rightarrow 0) = \frac{|\Psi(z\rightarrow horizon)|^{2}}{|\Psi(z\rightarrow boundary)|^{2}} - \frac{1}{i\omega}\frac{h'}{h}$. Now using \cite{Horowitz:2009ij}
$\Psi(z) = e^{-i \omega z} + R e^{i \omega z}$ for $ z >0 $ and $\Psi(z) = T e^{-i \omega z} $ near the horizon, we get $\sigma = \frac{T^{2}}{(1+ R)^{2}}$. Using $ T^{2} + R^{2} = 1$, we reach at $\sigma = \frac{1- R}{1+ R}$ .}. 
As we approach horizon we expect that reflection coefficient should decrease and hence $ \sigma_{Horizon}\geq \sigma_{Boundary}$ (as shown in the text).
 It will be interesting to explore further the Schr\"{o}edinger potential in detail, and extract out conductivity away from zero frequency limit.
\begin{itemize}
 \item {\textbf{Extremal case:}} In the inner region (close to horizon) for extremal black hole one finds 
(see sec.7)
\begin{equation}
\frac{d^2}{d\eta^2}\phi_i(\eta)+\phi_i(\eta)-\frac{a}{\eta^2}\phi_i(\eta)=0.\nonumber\\
\end{equation} Since $\eta$ contains a factor of $ \omega$, we define $x = \omega \eta$.
In this coordinate we get
\begin{equation}
\frac{d^2}{dx^2}\phi_i(x)+\omega^{2}\phi_i(x)-\frac{a}{x^2}\phi_i(x)=0.\nonumber\\
\end{equation}
Hence the  potential is given by
\begin{equation}
V_{Near horizon}(x)=\frac{a}{x^2} \label{extremalpotential}
\end{equation}
 At the horizon
\begin{equation}
\lim_{x \rightarrow \infty} V(x)=0 , 
\end{equation}
which is in agreement with non-extremal case (where $x \to \infty$, corresponds to horizon). Following argument in \cite{Horowitz:2009ij} we conclude at extremality\footnote{Author is thankful to referee for pointing this out.} $Re(\sigma) = \omega^{a} $. Note that $ a = 2 $, for all the examples considered in the text. 
\end{itemize}

% \begin{figure}[!]
%  \begin{minipage}[t]{6cm}
%  \vspace{-10pt}
%  \centerline{\hspace{6.3mm}
%  \rotatebox{0}{\epsfxsize=7cm\epsfbox{plot3.eps}}}
%  \hspace{3.3cm}\caption[]{Variation of inner region potential at extremality with respect to radial coordinate $y(=\frac{1}{\eta})$.Horizon at $y = 0$.Observe $V > 0$.} 
%  \protect\label{fig3}
%  \end{minipage}
%  \hfill
%  \begin{minipage}[t]{6cm}
%  \vspace{-10pt}
%  \centerline{\hspace{11.3mm}
%  \rotatebox{0}{\epsfxsize=7cm\epsfbox{plot4.eps}}}
%  \hspace{3.3cm}\caption[]{Variation of outer region potential at extremality with respect to radial coordinate $x(=1-u)$.Horizon at $ x = 0$. Observe $V > 0$.}
%  \protect\label{fig4}
%  \end{minipage}
%  \end{figure}

\section{Discussion}
Some of the issues that are studied and worth exploring are
\begin{itemize}
 \item Using our general results on perturbation equations we found out an expression for electrical conductivity for general background and they are different than $\mu=0$ cases. Effect of turning on chemical potential can be thought of as turning on effective interaction in the back ground.  Note that conductivity satisfies $\sigma_{Horizon}>\sigma_{Boundary}$. More precisely variation of $\sigma$ with $u$ is given by 
\begin{equation}
\sigma(u) = \frac{1}{8\pi G_{d+1}}G_{II}g^{xx}\sqrt{\frac{-g}{g_{uu}g_{tt}}}\Big[\frac{n H_{i}(u)}{n H_{i}+(H_{i}F-H_{i}^{'})(1-u^{n})}\Big]^{2}.
\end{equation}
At the horizon one gets
\begin{equation}
\sigma =\frac{1}{8\pi G_{d+1}}G_{II}g^{xx}\sqrt{\frac{-g}{g_{uu}g_{tt}}}.
\end{equation}
One can arrive at the form of the conductivity at the horizon using membrane paradigm arguments as developed in \cite{Iqbal:2008by}. 
\end{itemize}

 \begin{itemize}
  \item In the extremal case it is shown that at the horizon potential remains finite and following \cite{Horowitz:2009ij} it may be concluded that conductivity is zero in this case.
We also observe that dimension of dual operator can be given purely in terms of geometry and gauge fields. To arrive at that form we only have assumed double pole structure of the metric. So we expect result is independent of back ground . In all the examples considered, the dimension of dual operator is always $2$.
 \end{itemize}
\begin{itemize}
 \item It is also shown that thermal conductivity and viscosity obeys Wiedemann-Franz like law even in the presence of multiple chemical potential. It would be interesting to find out whether numbers $8 \pi^2, 32 \pi^2$ and $ 2\pi^2$, can be fixed in terms of central charges of dual gauge theories \cite{Jain:2009bi}. This will give us important information about universal behavior of dual gauge theories. It was shown in \cite{Kovtun:2008kx}, that conductivity in the case of vanishing chemical potential can be determined exactly in terms of central charges. It is desirable to investigate whether
electrical conductivity can also be fixed in terms of central charges in the presence of chemical potentials .
\end{itemize}
\begin{itemize}
 \item It would be interesting to study  $\kappa_T$ and $\frac{\kappa_T(\mu_{1}^{2}+\mu_{2}^{2}+\mu_{3}^{2})}{\eta T}$  at extremality. Here one needs to be extremely careful about properly defining various quantities. But naively if we apply results obtained at extremality (take for example $ AdS_{4}$ Reissner-Nordstr\"{o}m background, see \cite{Edalati:2009bi} for details)
\begin{equation}
\eta \sim \mu^{2}, ~~~~~\sigma \sim \Big(\frac{\omega}{\mu}\Big)^2.  
\end{equation} 
Then  taking\footnote{$\epsilon= 4\frac{r_{0}^3}{k_{4}^2 L^{4}},~~~ P =\frac{\epsilon}{2},~~~\mu=\frac{\sqrt{3} r_0}{L^2},~~\rho = 2 \sqrt{3} \frac{r_{0}^2}{k_{4}^2 L^2} $, see \cite{Edalati:2009bi}.}  $\epsilon + P = \rho \mu $, we obtain
\begin{equation}
\kappa_T \sim \frac{\omega^{2}}{T}, ~~~~  \frac{\kappa_T \mu^2}{\eta T} \sim \Big(\frac{\omega}{T}\Big)^{2}.
\end{equation} 
\end{itemize}

\section*{Acknowledgments}
It is pleasure to thank Somen Bhattacharjee, Sudipta Mukherji, Balram Rai, Sayan Chakrabarti, Anirban Basu, Sankhadeep Chakraborty , Binata Panda, Ambresh Shivaji, Souvik Banerjee for useful discussions. Special thanks to Balram Rai for discussion on chemical potential  and to Sayan Charkrabarti, Sankhadeep Chakraborty, Binata Panda and in particular Sudipta Mukherji about improving the manuscript. Author is also thankful to referee for pointing out mistakes in the form of Schr\"{o}edinger potential. Author would like to mention that he misses Prof. Alok Kumar and his constant encouraging words.  
\renewcommand{\thesection}{\Alph{section}} 
\setcounter{section}{0} 

\section{Appendix}
Here we collect all the geometrical as well as thermodynamical quantities required for computation done in the text (see \cite{Behrndt:1998jd},\cite{Jain:2009uj},\cite{Maeda:2008hn}, \cite{Gubser:1998jb} for details).

\subsection{Five dimensional black hole}
In this subsection, we collect all the relevant information's for five dimensional R-charged black hole.
 Lagrangian is
\begin{equation}
 {{\cal L}\over
 \sqrt{-g} }=
 R  - {1\over 4} G_{ij} F_{\mu\nu}^i 
 F^{\mu \nu\, j} +.....
\label{lagrangian}
\end{equation}
Where
$$
G_{ij} = {L^{2}\over 2} \mbox{diag} \left[ (X^1)^{-2}, \, (X^2)^{-2}, \,(X^3)^{-2}
\right]\,.
$$
Metric and gauge fields are
\begin{equation}
ds^2_5 = - {\cal H}^{-2/3}{(\pi T_0 L)^2 \over u}\,f \, dt^2 
+  {\cal H}^{1/3}{(\pi T_0 L)^2 \over u}\, \left( dx^2 + dy^2 + dz^2\right)
+ {\cal H}^{1/3}{L^2 \over 4 f u^2} du^2\,,
\label{metric_u_3}
\end{equation}
\begin{equation}
f(u) = {\cal H} (u) - u^2 \prod\limits_{i=1}^3 (1+\kappa_i)\,, 
\;\;\;\;\; H_i = 1 + \kappa_i u \,, \;\;\;\;\; 
{\cal H}=\displaystyle\prod_{i=1}^3 H_i 
\label{identif}
\end{equation}
\begin{equation}
X^i = {{\cal H}^{1/3}\over H_i(u)} \,, \qquad 
A^i_t = { \pi T_0 \sqrt{2k_i(1+k_1)(1+k2)(1+k3)} u\over  H_i(u)}\,,\qquad  
\label{scal_gauge_u_3}
\end{equation}
Viscosity and various thermodynamical quantities are given by
\begin{equation}
T_H = 
{2 + \kappa_1 + \kappa_2 + \kappa_3 - \kappa_1 \kappa_2  \kappa_3\over 
2\sqrt{(1+\kappa_1)(1+\kappa_2) (1+\kappa_3)}}\, T_0\,.
\end{equation}
\begin{equation}
s =  {\pi^2 N^2 T_0^3 \over 2} 
 \prod\limits_{i=1}^3 (1+\kappa_i)^{1/2}  \, , 
\label{entropy_density}
\end{equation}
\begin{equation}
\eta =  {\pi N^2 T_0^3 \over 8} 
 \prod\limits_{i=1}^3 (1+\kappa_i)^{1/2}  \, , 
\label{eta}
\end{equation}
\begin{eqnarray}
\varepsilon &=& { 3 \pi^2 N^2 T_0^4 \over 8}   \prod\limits_{i=1}^3
 (1+\kappa_i)\,,\label{energy_density} \\
  P &=&  {\pi^2 N^2 T_0^4 \over 8}   \prod\limits_{i=1}^3
 (1+\kappa_i)\,. 
\label{pressure}
\end{eqnarray}
The densities of physical charges are 
\begin{equation}
\rho_i = 
{\pi N^2 T_0^3\over 8} \sqrt{2 \kappa_i}  \prod\limits_{l=1}^3
 (1+\kappa_l)^{1/2}\,.  
\end{equation}
The chemical potentials conjugated to $\rho_i$ are defined as
\begin{equation}
\mu_i = A_t^i (u)\Biggl|_{u=1} = {\pi T_0 \sqrt{2 \kappa_i}\over (1+\kappa_i)}
 \prod\limits_{l=1}^3
 (1+\kappa_l)^{1/2}\,.  
\end{equation}
and 

\begin{equation}
 \frac{1}{16\pi G_{5}} =\frac{N_c^{2}}{16 \pi^{2}L^{3}}.
\end{equation}
\subsection{Four dimensional black hole}
Metric and gauge fields in this case are
\begin{subequations}
\bea
ds_4^2
  &=& \frac{16(\pi T_0 L)^2}{9u^2}{\cal H}^{1/2}
    \left( - \frac{f}{{\cal H}} dt^2 + dx^2 + dz^2 \right)
  + \frac{L^2}{f u^2}{\cal H}^{1/2}~du^2~, \\ 
A_{t}^{i}  &=& \frac{4}{3} \pi T_0 \sqrt{2 \kappa_i\prod\limits_{i=1}^4
 (1+\kappa_i)}~\frac{u}{H_i}~, \\  
H_i &=& 1 + k_i u~, \\
{\cal H}&=&\prod\limits_{i=1}^4 H_i, \\
f &=& {\cal H}-\prod\limits_{i=1}^4
 (1+\kappa_i)u^{3} ,
\eea
\end{subequations}
Thermodynamic quantities are given by 

\begin{subequations}
\begin{align}
  & \epsilon
  = \sqrt{2}\, \pi^2\, \left( \frac{2}{3} \right)^4\,
    N_c^{3/2}\, T_0^3~\prod\limits_{i=1}^4(1 + \kappa_i)~,
& & P
  = \frac{\sqrt{2}\, \pi^2}{3}\, \left( \frac{2}{3} \right)^3\,
    N_c^{3/2}\, T_0^3~\prod\limits_{i=1}^4(1 + \kappa_i)~,
\label{eq:energy_density-P-M2} \\
  & s  =\sqrt{2}\pi^2 \left( \frac{2}{3} \right)^3 N_c^{3/2}\, T_0^2~\prod\limits_{i=1}^4\sqrt{1 + \kappa_i}~,
& & T
  = \begin{scriptsize}\frac{T_0\left(3+\sum\limits_{j=1}^4 k_i+\sum\limits_{j>i,i,j=1}^4k_i k_j-\prod\limits_{i=1}^4k_i \right)}{3\sqrt{\prod\limits_{i=1}^4(1 + \kappa_i)}}                                                                                                                                                              \end{scriptsize}
\label{eq:entropy_density-T_H-M2} \\
  & \rho_i
  = \sqrt{2}\, \pi\, \left( \frac{1}{3} \right)^3\,
    N_c^{3/2}\, T_0^2~\sqrt{2\, k_i\prod\limits_{j=1}^4(1 + \kappa_j) }~,
& & \mu_i
  = \frac{4 \pi\, T_0}{3}\, \frac{1}{1+k_i}\sqrt{ 2\, \kappa_i \prod\limits_{i=1}^4(1 + \kappa_i) }~,
\label{eq:charge_density-chemical_potential-M2}
\end{align}
\end{subequations}
Other relevant expressions are
\begin{equation}
\eta=\frac{1}{4}\sqrt{2}\, \pi\, \left( \frac{2}{3} \right)^3\,
    N_c^{3/2}\, T_0^2~\prod\limits_{i=1}^4\sqrt{1 + \kappa_i}
\label{eta4}
\end{equation}
$$
G_{ij} = {L^{2}\over 2} \mbox{diag} \left[ (X^1)^{-2}, \, (X^2)^{-2}, \,(X^3)^{-2}, \,(X^4)^{-2}
\right]\,.
$$
\begin{equation}
X^i = {{\cal H}^{1/4}\over H_i(u)} 
\end{equation}
\begin{equation}
 \frac{1}{16\pi G_{4}} =\frac{N_c^{\frac{3}{2}}}{24\sqrt{2}L^{2}}.
\end{equation}
\subsection{Seven dimensional black hole}
\begin{subequations}
\bea
ds_7^2
  &=& \frac{4(\pi T_0 L)^2}{9u}{\cal H}^{1/5}
    \left( - \frac{f}{{\cal H}} dt^2 + dx_1^2 + \cdots + dx_4^2 + dz^2 \right)
  + \frac{L^2}{4 f u^2}{\cal H}^{1/5}~du^2~, \\ 
A_{t}  &=& \frac{2}{3} \pi T_0 \sqrt{2 \kappa_i\prod\limits_{i=1}^2(1+\kappa_i)}~\frac{u^2}{H_i}~, \\  
H_i &=& 1 + \kappa_i u^2~, \\
{\cal H}&=&\prod\limits_{i=1}^2 H_i, \\
f &=& {\cal H}-\prod\limits_{i=1}^2(1+\kappa_i)u^{3}~,
\eea
\end{subequations}
Thermodynamic quantities are given by 
\begin{subequations}
\begin{align}
  & \epsilon
  = \frac{5\, \pi^3}{2}\, \left( \frac{2}{3} \right)^7\,
    N_c^3\, T_0^6~\prod\limits_{i=1}^2( 1 + \kappa_i )~,
& & P
  = \frac{\pi^3}{2}\, \left( \frac{2}{3} \right)^7\,
    N_c^3\, T_0^6~\prod\limits_{i=1}^2( 1 + \kappa_i )~,
\label{eq:energy_density-P-M5} \\
  & s  = 3\, \pi^3\, \left( \frac{2}{3} \right)^7\,
    N_c^3\, T_0^5~\sqrt{\prod\limits_{i=1}^2(1 + \kappa_i)}~,
& & T
  = \frac{T_0\left( 3 +\kappa_1+\kappa_2- \kappa_1\kappa_2 \right)}{3\sqrt{\prod\limits_{i=1}^2(1 + \kappa_i)}}~,
\label{eq:entropy_density-T_H-M5} \\
  & \rho_i
  = \pi^2\, \left( \frac{2}{3} \right)^6\,
    N_c^3\, T_0^5~\sqrt{2\, \kappa_i\, \prod\limits_{i=1}^2( 1 + \kappa_i )}~,
& & \mu_i
  = \frac{2 \pi\, T_0}{3(1+\kappa_i)}\, \sqrt{ 2\, \kappa_i\prod\limits_{i=1}^2( 1 + \kappa_i )} ~,
\label{eq:charge_density-chemical_potential-M5}
\end{align}
\end{subequations}
Other relevant results are
\begin{equation}
\eta=\frac{3}{4}\, \pi^2\, \left( \frac{2}{3} \right)^7\,
    N_c^3\, T_0^5~\sqrt{\prod\limits_{i=1}^2(1 + \kappa_i)}
\label{eta7}
\end{equation}
$$
G_{ij} = {L^{2}\over 2} \mbox{diag} \left[ (X^1)^{-2}, \, (X^2)^{-2}\right]\,.
$$
\begin{equation}
X^i = {{\cal H}^{2/5}\over H_i(u)} 
\end{equation}
\begin{equation}
 \frac{1}{16\pi G_{7}} =\frac{ N_c^{3}}{6 \pi^{3} L^{5}},
\end{equation}
\section{Blackholes at extremality}
Above black holes at extremality was constructed in \cite{Lu:2009gj}. 
Take 
\begin{equation}
\bar g_{tt} = - f(u) A_1(u) , ~~\bar g_{uu} = A_2(u) f^{-1}(u)  ,~f(u)=(1-u)^2 V(u).
\end{equation}
Here we just give relevant information about $f$.
\begin{center}
\begin{tabular}{l*{2}{c}r}
 Dimension         &  Extremality condition   &$V(u)$ \\
\hline
  5           &$2 + \kappa_1 + \kappa_2 + \kappa_3 - \kappa_1 \kappa_2  \kappa_3=0$&$(1 + \kappa_1 \kappa_2 \kappa_3 u)$ \\
  4          &$ 3 +\sum\limits_{j=1}^4 k_i+\sum	\limits_{i<j,i,j=1}^4 k_i k_j-\prod\limits_{i=1}^4 k_i =0$ &$(1 + (2 +\sum\limits_{j=1}^4 k_i ) u +\prod\limits_{i=1}^4 k_i  u^2)$ \\
  7          &$3 +\kappa_1+\kappa_2- \kappa_1\kappa_2=0$&$(1 + 2 u + \kappa_1 \kappa_2 u^2) $ \\
\end{tabular}             \end{center}

\end{document}